\documentclass[10pt]{article}

\usepackage[OE]{express}
\usepackage{hyperref}

\usepackage{amsmath}

\usepackage{cite}

\begin{document}

\title{Polarization dependence of transverse photo-induced voltage in gold thin film with random nanoholes}
\author{Marjan Akbari and Teruya Ishihara $^{*}$}
\address{Department of Physics, Tohoku University, Sendai, 980-8578, Japan}

\email{\authormark{*}t-ishihara@m.tohoku.ac.jp}

\begin{abstract} Transverse photo-induced voltage ({\small TPIV}) in 25nm-thick Au film with random holes with 100 nm in diameter is measured for linearly, circularly and elliptically polarized light. By rotating the major axis of ellipse of the light, {\small TPIV} exhibits specific pattern depending on polarization. The experimental results are readily reproduced by assuming that the angular momentum transfer from the light beam to the film is responsible for {\small TPIV}. A novel ellipticity meter is proposed based on this mechanism.
\end{abstract}


\section{Introduction}

When a circularly polarized light beam is irradiated on gyrotropic crystals, voltage depending on the sense of circular polarization is generated, which is referred to as a circular photogalvanic effect \cite{IvchenkoPikus1978}.  Similar effect is observed when the beam is incident obliquely on a non gyrotropic material: a transverse voltage can be generated perpendicular to the incident plane \cite{Ivchenko2002}. The effect is believed to be based on the conversion of angular momentum of photon to the sample and even incident beam with linear polarized light except for s- and p- polarized light may generate it. \\
\indent
The phenomena have been observed in many systems such as quantum wells \cite{Ganichev2001, Shalygin2006}, metallic photonic crystal slabs \cite{Hatano2009}, Ag-Pd film \cite{Mikheev2011}, single walled carbon nanotube film \cite{Mikheev2012A}, graphene \cite{Jiang2014} and spongy nanoporous gold thin film \cite{Akbari2015}. The phenomenon has been ascribed to different mechanisms including circular photogalvanic effect, circular photon drag effect \cite{Jiang2014} and circular ac Hall effect \cite{Karch2010}. \\ 
\indent
Thin film of gold with random nanohole created by nanosphere lithography ({\small NSL}) is one of these random nanostructures that supports both localized and propagating surface plasmon resonances \cite{chang2005}. Besides, enhanced optical transmission through subwavelength hole structures has attracted interest to optical phenomena and applications of nanoholes in metal films \cite{Rindzevicius2006}. Because of these features, this nanostructure has a good potential to serve as a plasmon nanohole photodetectors in visible range and various optoelectronic devices and applications in various fields such as optics, biological sensing and photovoltaic \cite{Ohno2016}. Random and ordered nanostructures have been fabricated \cite{Li2013} and among them, two dimensional plasmonic nanostructures made by {\small NSL} are particularly suitable for commercialization because this is an easy way for low cost manufacturing method \cite{Zheng2015}. \\
\indent
In our previous paper \cite{Akbari2015}, we reported that a spongy nanoporous gold thin film gives fairly large transverse photo-induced voltage ({\small TPIV}) as well as longitudinal photo-induced voltage ({\small LPIV}) signals compared to flat Au film. In this paper we investigate polarization dependence of {\small TPIV} in thin film of gold with random nanohole and compare it with the numerical calculation based on optical angular momentum transfer to the sample. \\
\indent
Here we report transverse photo-induced voltage generated by oblique incident laser light in visible wavelengths in gold thin film with random nanoholes with size of 100 nm fabricated by nanosphere lithography technique. Sign and amplitude of the voltage depend on the sense of polarization and ellipticity of the laser radiation. Calculations show that transferred angular momentum from photon to the sample upon reflection and transmission is responsible for this voltage.\\
\indent
Furthermore we propose an ellipticity meter based on this mechanism. Although ellipticity dependent photocurrent has been discussed in other systems, it is the first proposal of such a type of device to determine ellipticity to the author's best knowledge best knowledge. This device can be fabricated from only one 25 nm layer of this material and used to distinguish linearly, circularly and elliptically polarized light, which is compared to the linear polarization analyzer based on nanographite \cite{Mikheev2012B}.\\

\begin{figure}[t]
\begin{center}
\includegraphics[trim = 0mm 50mm 0mm 40mm, clip, width=\linewidth]{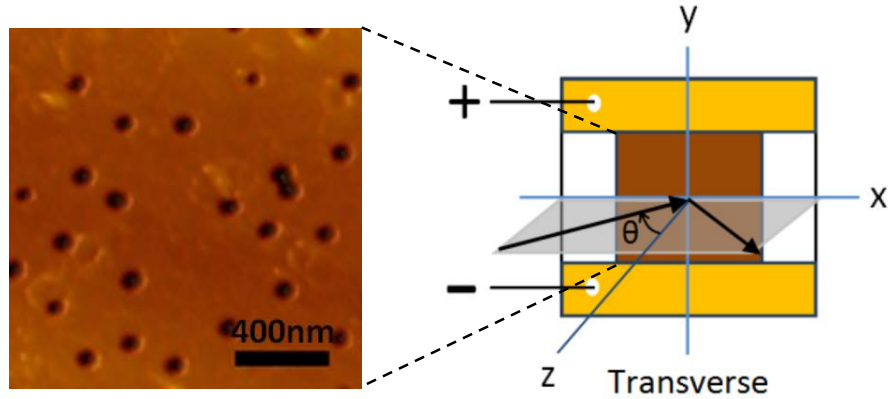}
\caption{\label{Sample1} AFM image of gold thin film with random nanoholes and transverse configuration. The arrows show definition of the positive incidence angle ($\theta$).}
\end{center}
\end{figure}

\section{Experimental}
\subsection{Sample and experimental setup}

{\bf NSL technique.} For fabrication of gold thin film with random nanoholes we used {\small NSL} technique. In this method, 15 mm $\times$ 15 mm glass substrate was cleaned by successive sonication in acetone, ethyl alcohol and distilled water for 5 min. We used polystyrene ({\small PS}) colloids (diameter 100 nm) to make nanoholes. For attachment of the {\small PS} colloids to the substrate a solution of aluminum chloride hydroxide (${\small  AlCl}_{3}.6{\small H}_{2}{\small O}$) \cite{Junesch2014} was deposited on the substrate by sonication for 5 min. For preparing this solution, a mixture of 95 wt.$\%$ of distilled water and 5 wt.$\%$ of ${\small AlCl}_{3}.6{\small H}_{2}{\small O}$ was provided. \\
\indent
Substrate was removed and rinsed in distilled water for 2 min. Then it was blow dried with $N_{2}$ gas. One drop of {\small PS}  colloids was coated on the glass by putting it for 1min on the prepared glass, followed by distilled water rinse to remove extra {\small PS} colloids. Then wet glass was immersed in the $130^\circ$ C hot ethylene glycol to fix the {\small PS} spheres. Then substrate was rinsed with distilled water and blow dried with ${\small N}_{2}$ gas.\\

\noindent
{\bf Metal deposition.} On the substrate coated with {\small PS}  colloid particles as a mask, a 25 nm layer of gold was deposited by sputtering after a 3 nm layer of chromium was deposited as adhesion layer. Then {\small PS}  spheres were removed by tape stripping. The result is a 25 nm thick gold film with nanohole size of 100 nm. Film was trimmed with a focused laser beam into dimensions of 5 mm $\times$ 5 mm. After the attachment of the electrodes, the substrate was fixed on a 20 mm $\times$ 20 mm printed circuit board ({\small PCB}). The typical Ohmic resistance of the films was 100 $\Omega$.\\

\noindent
{\bf Configuration.} In our experiment, we employed transverse configuration. In this configuration, two electrodes were oriented parallel to the incidence plane and the voltage along y-axis was measured. Positive incident angle is shown by the arrow (Fig. \ref{Sample1}). (In longitudinal configuration, the electrodes are oriented perpendicular to this plane and the voltage along x-axis was measured). Note that our definition of configuration differs from the one adopted in some of the references\cite{Mikheev2012A}. \\

\begin{figure}[t]
\begin{center}
\includegraphics[trim = 45mm 35mm 40mm 20mm, clip, width=\linewidth]{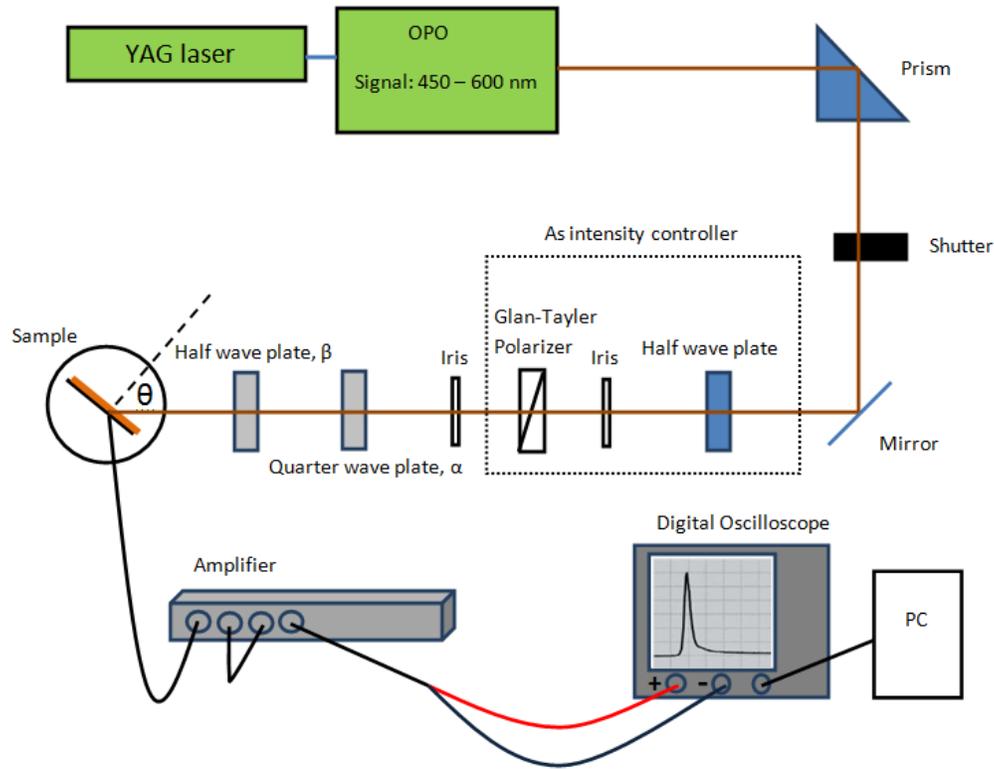}

\caption{\label{Configs}Setup for measuring {\small TPIV} with different polarized laser light.}
\end{center}
\end{figure}
\noindent
{\bf Setup.} A schematic diagram of the experimental setup is shown in Fig. \ref{Configs}. Light with wavelengths of 450-600 nm was generated by an optical parametric oscillator ({\small OPO}) excited by a tripled  {\small Nd:YAG} laser. The laser pulse width and repetition rate were 7 ns and 10 Hz, respectively. The polarization of light was controlled with an achromatic half wave plate ({\small HWP}) and quarter wave plate ({\small QWP}), which are valid for wavelength between 400 and 800 nm. The beam size was chosen so that it covers the sample at the normal incidence. Area around the film was  masked so that light does not impinge on the electrodes for larger incidence angles and we make sure that the voltage is generated only by the porous gold area. \\
\indent
Intensity of incident light was kept constant about 4 MW/cm$^2$ at normal incidence for all wavelengths (threshold of the laser damage intensity was 24 MW/cm$^2$) using a {\small HWP} on a rotational stage and a Glan-Tayler polarizer passing the horizontally polarized light. Laser pulse energy for this intensity was 0.3 mJ. We measured {\small TPIV} at a fixed delay with a digital oscilloscope (50 $\Omega$ input impedance and pass band of 100 MHz) triggered by {\small Q}-switch of the laser. \\
\indent
We found that the {\small TPIV} signal follows the light pulse profile. Before feeding voltage to the oscilloscope, an amplifier with a gain of 125 was used. All the voltages referred in this paper, however, are not the amplified ones but the ones generated at the sample. For reducing pulse to pulse fluctuations in the {\small TPIV} signals, they were averaged for 32 pulses. Wavelength and {\small HWP} setting and data acquisition are controlled with a LabVIEW program.\\
\indent
In this paper, we adopt the definition for the circularly polarized light in the field of quantum physics: Looking at the light along its propagations direction, ${\bf E}$ field of the {\small RCP} light rotates clockwise. 

\subsection{Wavelength dependence}  

\begin{figure}[t]
\footnotesize
\includegraphics[width=\linewidth]{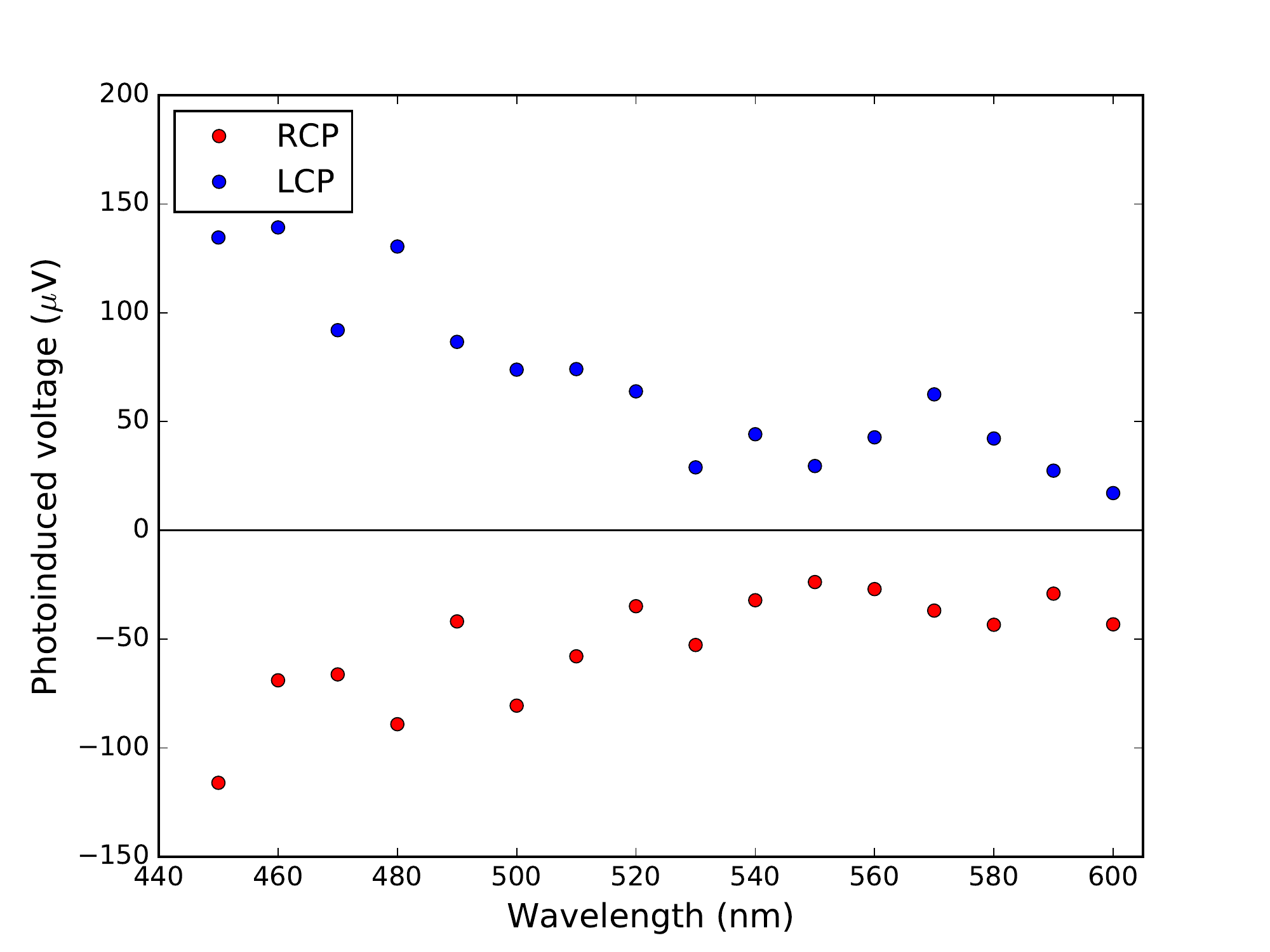}  

\caption{\label{TLWRL} Wavelength resolved {\small TPIV} for $+50^\circ$ incidence angle with {\small RCP} and {\small LCP} light measured in gold film with random nanoholes. }
\end{figure}

We measured {\small TPIV} along y-axis of the gold thin film in transverse configuration with linearly, circularly and elliptically polarized light. For this experiment the {\small HWP} just before the sample in  Fig. \ref{Configs} was removed. Figure \ref{TLWRL} shows dependence of the {\small TPIV} on the wavelength, for transverse configuration with right hand circularly polarized ({\small RCP}) and left hand circularly polarized ({\small LCP}) light at fixed incidence angle, $\theta$= $+50^\circ$. \\
\indent
As it is clear from this figure, intensity of voltage between 450 to 520 nm gradually decreases but for longer wavelengths it is almost the same. Sign of the {\small TPIV} does not change by changing wavelength in this range; for {\small LCP} it is always positive and for {\small RCP} always negative. Similar wavelength dependence is observed for linearly and elliptically polarized light. This behavior allows us to use this gold film with random nanoholes as a broadband ellipticity meter in this range.

\subsection{ {\small TPIV} for various polarization states with rotated orientation}

\begin{figure}[t]
\footnotesize
\includegraphics[trim = 15mm 10mm 10mm 10mm, clip, width=0.5\linewidth]{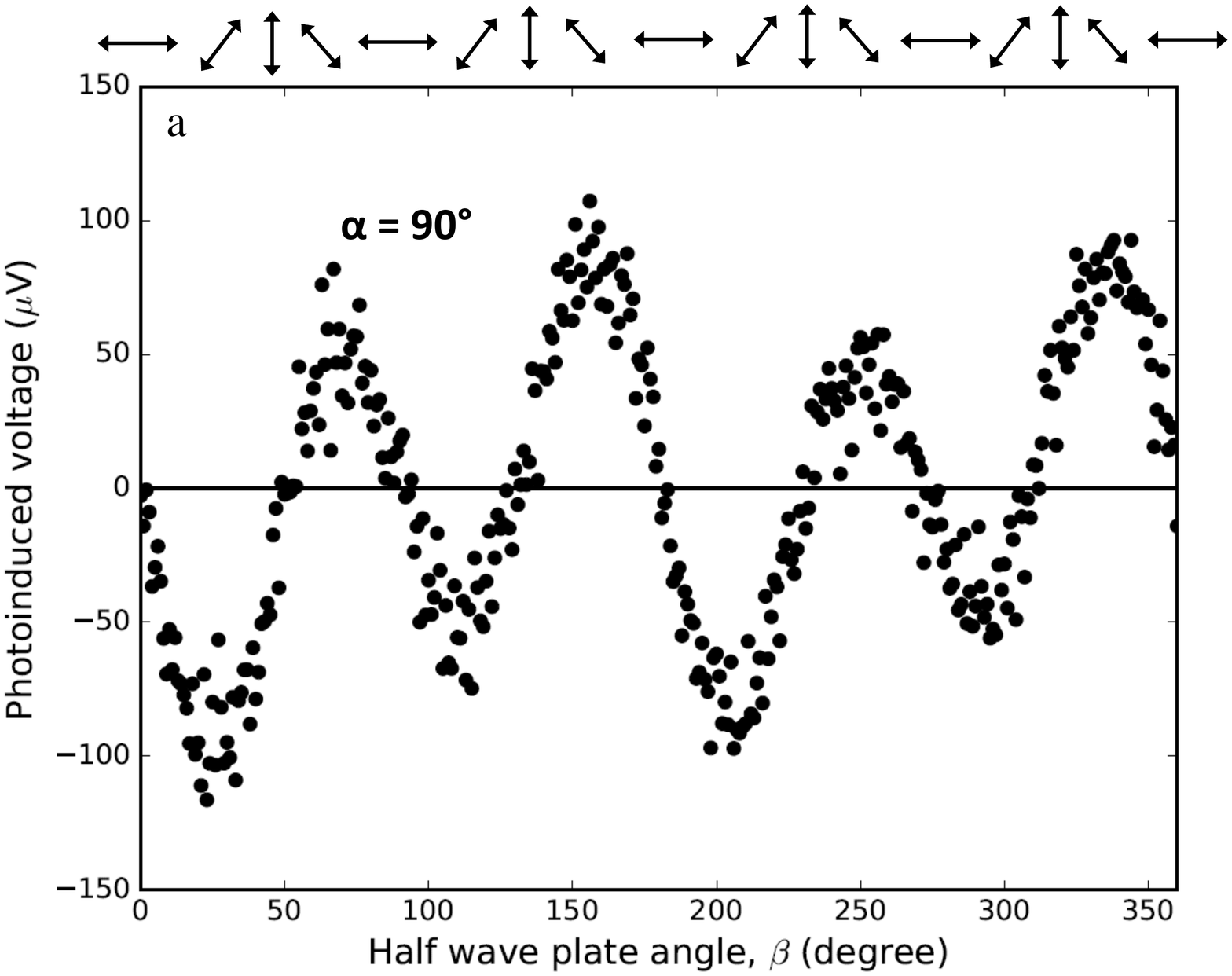} 
\includegraphics[trim = 15mm 10mm 10mm 10mm, clip, width=0.5\linewidth]{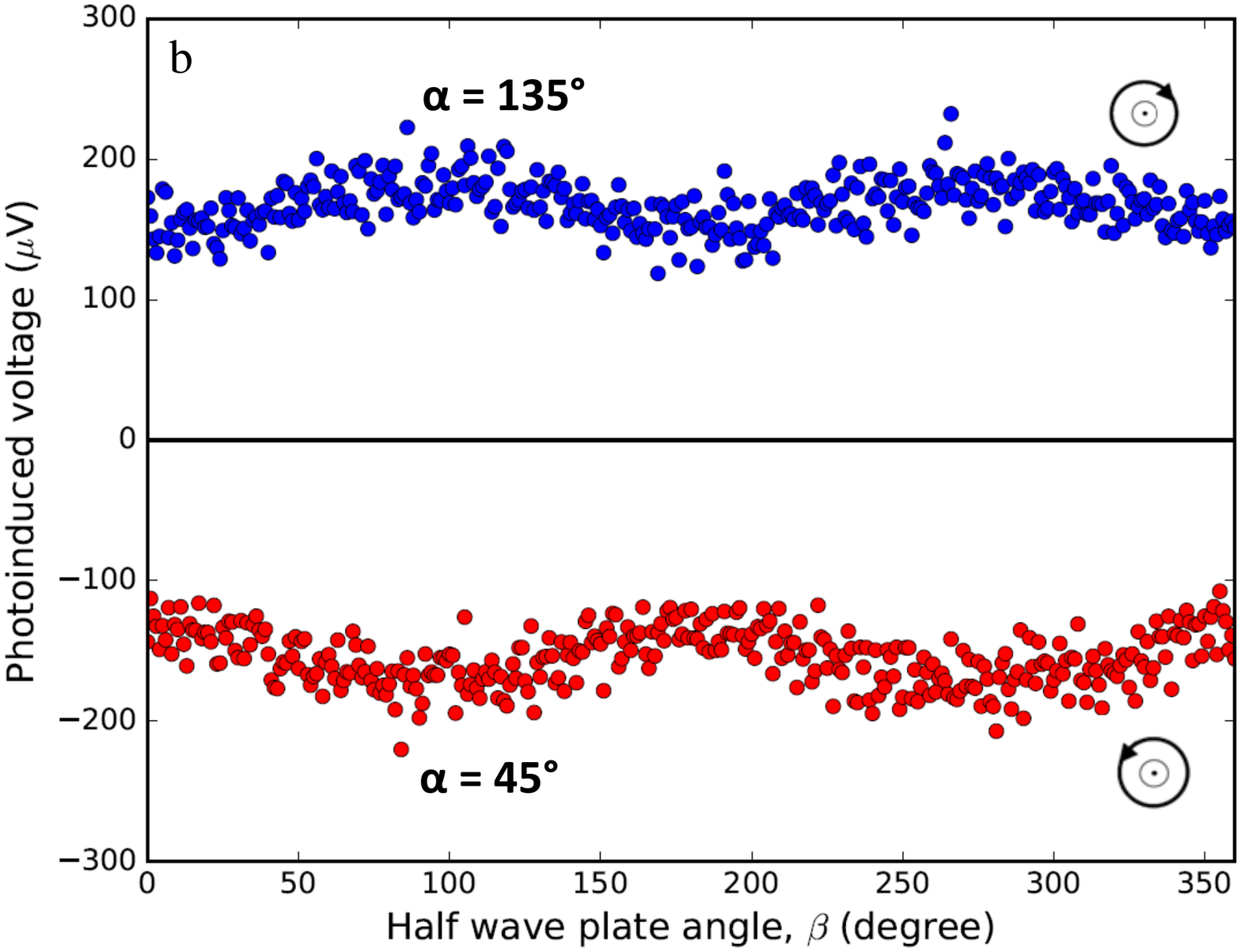} 
\caption{\label{lin}Measured {\small TPIV} for wavelength of 450 nm and $+50^\circ$ incidence angle in gold thin film with random nanoholes as a function of {\small HWP} angle. (a) for linearly polarized light (b) for circularly polarized light: Red dot: {\small RCP}, blue dot: {\small LCP}.}
\end{figure}

{\bf {\small TPIV} with linearly polarized light.} First we set the  {\small QWP} fast axis angle $\alpha$ to be  $90^\circ$, while the incident beam is polarized horizontally. In order to investigate orientation dependence of {\small TPIV}, we inserted {\small HWP} before the sample. Note that the sense of rotation of elliptically polarized light is inverted by this insertion. Thus {\small RCP} is converted to {\small LCP} and vice versa. The incidence angle, ${\theta}$, is fixed to be $+50^\circ$. \\
\indent
By rotating the {\small HWP}, direction of the fast axis rotates, therefore the orientation of the ${\bf E}$ field of the passing light through it changes and as we can see in top of the Fig. \ref{lin} all orientations of linear polarization are generated which includes s- polarization (vertical ${\bf E}$ field), p- polarization (horizontal ${\bf E}$ field) and other angles of orientation between vertical and horizontal polarizations. \\
\indent
As this graph shows sign and amplitude of the {\small TPIV} generated by the light depends on the orientation of the input light on the sample. For example, for vertical and horizontal polarizations, in the transverse configuration, {\small TPIV} cannot be generated; therefore photo voltage is zero for these polarizations. For other polarizations between vertical and horizontal, {\small TPIV} is generated and depending on the orientation of the ${\bf E}$ field, sign and amplitude of the {\small TPIV} changes periodically as we see in Fig. \ref{lin}.\\
\indent
As we can see in this graph, there is unequal response at equivalent angles, for example, for $\beta=22.5^\circ$ and $\beta=112.5^\circ$ for {\small HWP} angle which are supposed to create the same linear polarized light as is schematically shown on the top of the figure, photo response generated by the latter has a smaller amplitude compared to the former. In section 3 we will show that this unexpected non equality of the amplitudes for equivalent angles is related to the improper phase delay of the rotating {\small HWP}. We observed this effect for circular and elliptical polarized light as well.\\

\noindent
{\bf {\small TPIV} with circularly polarized light.} In the same setup, fixing angle between x-axis and fast axis of the  {\small QWP} at $\alpha=135^\circ$ and $\alpha=45^\circ$, we generated {\small RCP} and {\small LCP} light, which are converted by {\small HWP} to {\small LCP} and {\small RCP}, respectively. By rotating the {\small HWP}, direction of the fast axis rotates, but it should not affect the polarization state as the circular polarization does not have particular directions. As can be seen in Fig. \ref{lin}, the {\small HWP} angle dependence is obviously distinguished from linear polarization. The sign is constant and the amplitude is approximately the same. The sense of polarization rotation is clearly detected: positive for {\small LCP} and negative for {\small RCP}. The amplitude modulation is ascribed to the improper phase delay in {\small HWP}, as will be shown in section 3.\\

\begin{figure}[t]
\footnotesize

\includegraphics[trim = 15mm 15mm 20mm 5mm, clip, width=\linewidth]{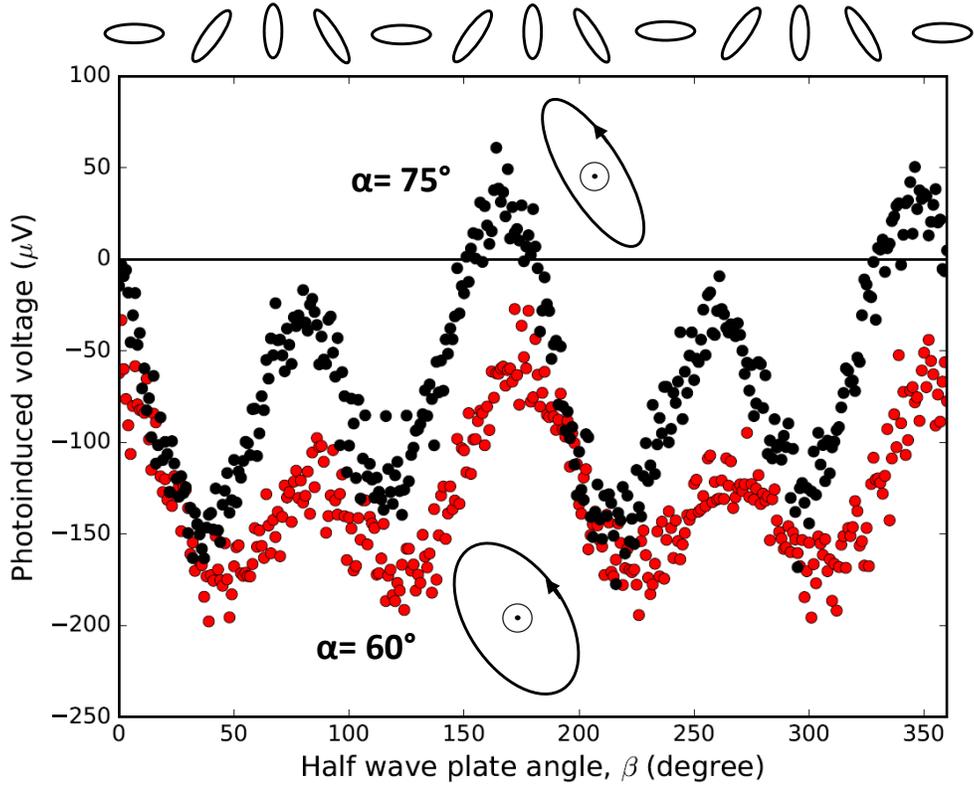}
\caption{\label{Ells} Measured {\small TPIV} as a function of {\small HWP} angle with 450 nm and $+50^\circ$ incidence angle in gold thin film with random nanoholes, Red dot: angle between x-axis and fast axis of the {\small QWP} is $60^\circ$, blue dot: angle between x-axis and fast axis of the  {\small QWP} is $75^\circ$. Ellipticities are different but both are right hand elliptical polarized.}
\end{figure}

\noindent
{\bf {\small TPIV} with elliptically polarized light ($\alpha= 60^\circ$ and $75^\circ$).} Increasing  $\alpha$ from $45^\circ$ to $90^\circ$,  we obtain right hand elliptically polarized light after the {\small HWP} before the sample. In Fig. \ref{Ells} the shapes of ellipse  are schematically shown (ellipse of light in the first case is wider).\\
\indent
The {\small TPIV} of the elliptically polarized light (red and black circles in Fig. \ref{Ells} correspond to $\alpha= 60^\circ$  and $75^\circ$, respectively) is between that of the {\small RCP} and linearly polarized light, because by changing direction of the fast axis of the {\small QWP}, ellipticity of the light changes from circular to linear. As we see in this figure, for $\alpha=75^\circ$ (black circles), for some of the angle of the {\small HWP}, {\small TPIV} changes its sign, which is somewhat similar to the response to the  linear polarized light.

\section{Discussion}

\begin{figure}[t]
\footnotesize
\includegraphics[trim = 15mm 10mm 30mm 25mm, clip, width=0.5\linewidth]{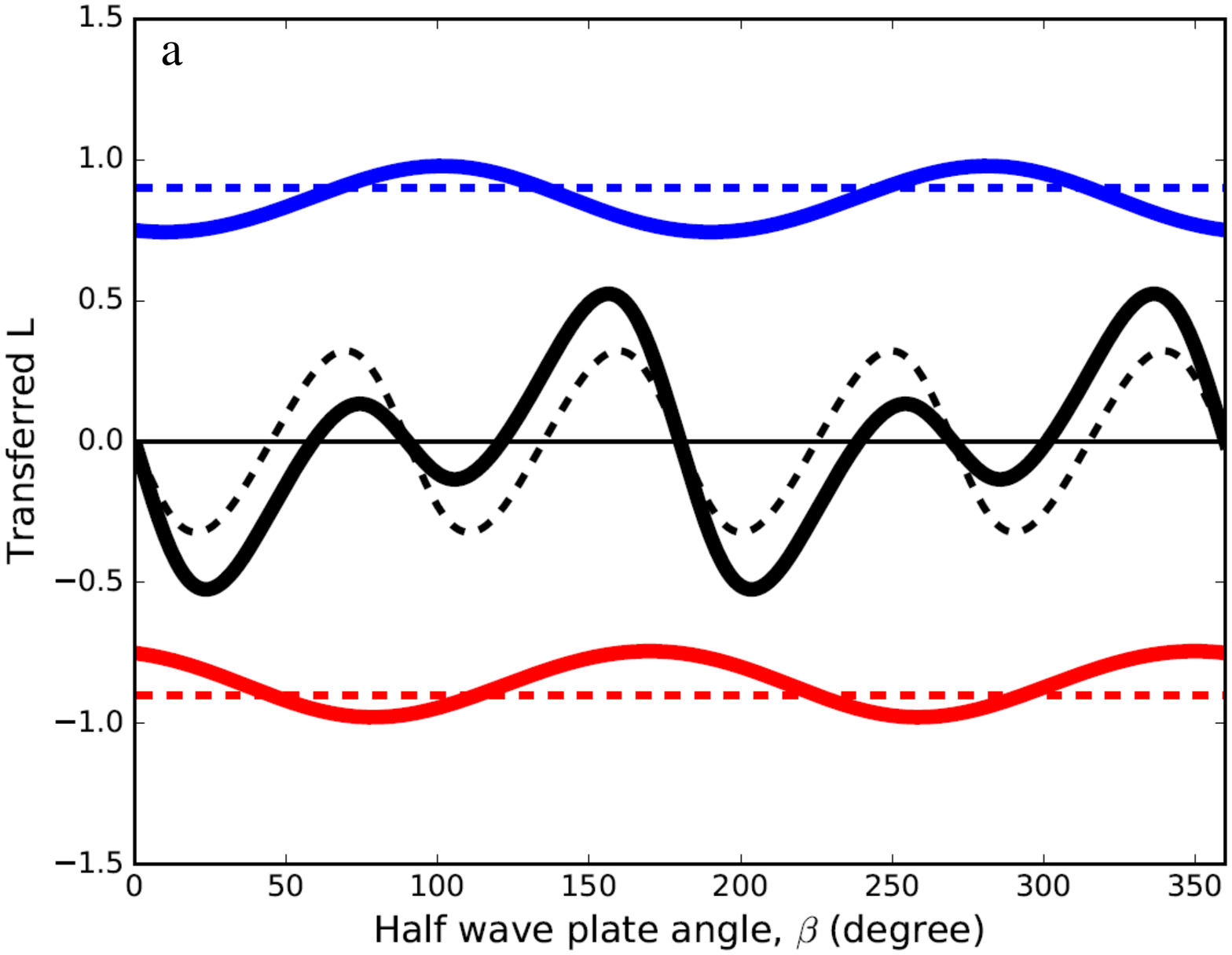} 
\includegraphics[trim = 13mm 10mm 30mm 25mm, clip, width=0.5\linewidth]{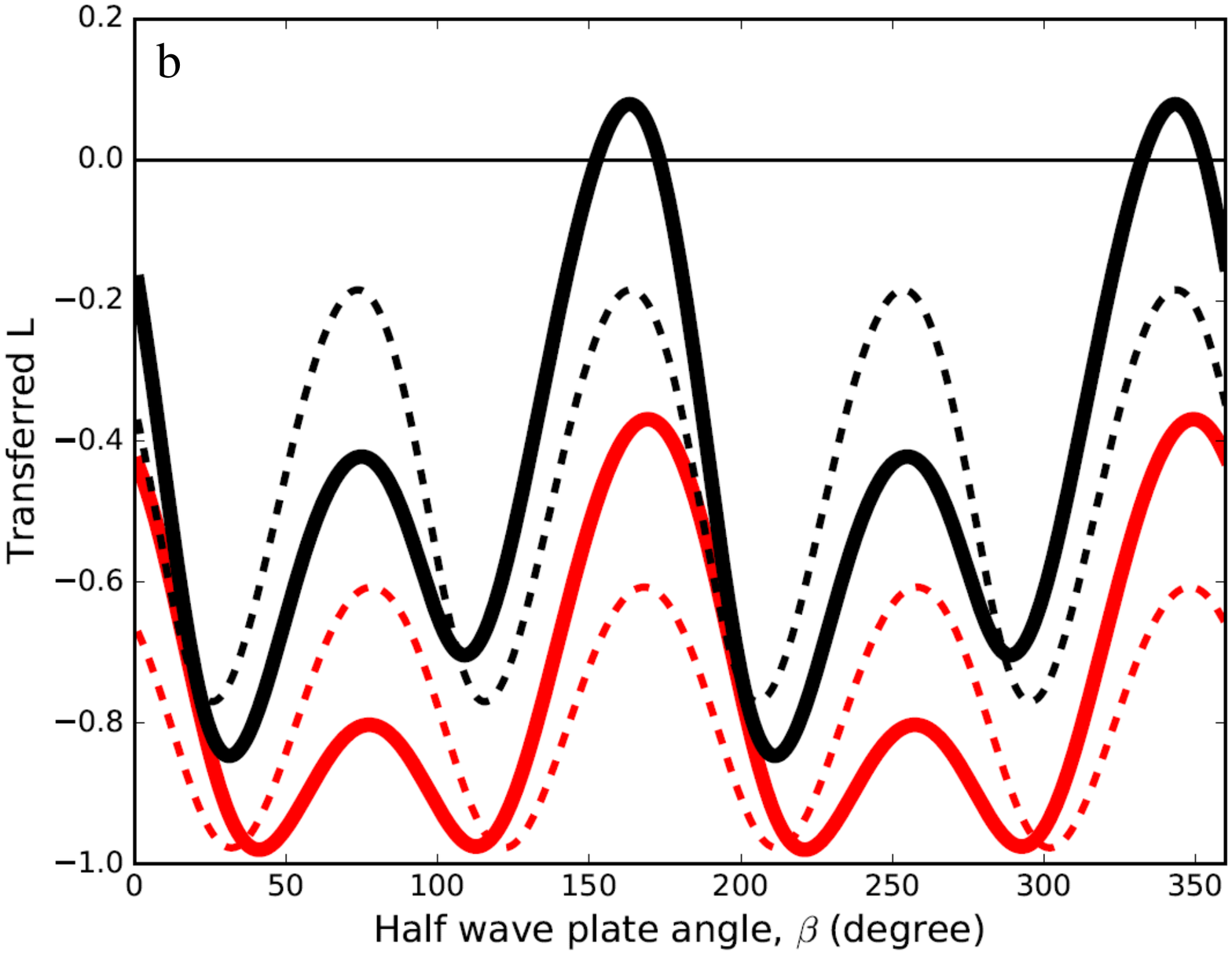} 

\caption{\label{All} Calculated $\Delta L_{Sample}$ as a function of {\small HWP} angle $\beta$ with $\theta = +50^\circ$ incidence angle: for linearly polarized light in gold thin film with disordered nanoholes, solid line: $\delta_{\small HWP}=0.9\pi$ and dashed line: $\delta_{\small HWP}=\pi$. a) black: $\alpha=0^\circ$, blue: $\alpha=135^\circ$,  red: $\alpha=45^\circ$. b) red: $\alpha=60^\circ$, black: $\alpha=75^\circ$.}
\end{figure}

{\bf Angular momentum transfer model.}
In order to understand the polarization dependence, we have calculated angular momentum transfer to the sample. Calculations were carried out by estimating of the transferred angular momentum from photon to the film upon reflection and transmission.\\
\indent
Using Jones matrices for p- polarized input light and  {\small QWP} we calculated output of the  {\small QWP} and its ellipticity, $\eta_{I}$. After the {\small QWP}, this light passes through the {\small HWP}. Using measured reflection and transmission coefficients, $ \tilde {r}_{p}=r_{p}exp({i\delta_{rp}})$,  $\tilde {r}_{s}={r}_{s}exp({i\delta_{rs}})$,  $\tilde {t}_{p}={t}_{p}exp({i\delta_{tp}})$ and $\tilde t_{s}=t_{s}exp({i\delta_{ts}})$ in Jones matrix for Fresnel equation, ellipticity of the reflected light and transmitted ligth can be calculated. Considering conservation of angular momentum in the system:

\begin{equation}
\begin{aligned}
\Delta L_\text{Sample}= L_\text{I}\cos{\theta}- ( L_\text{T}\cos{\theta}+ L_\text{R}\cos{(\pi -\theta)})\\
= L_\text{I}\cos{\theta}-  L_\text{T}\cos{\theta}+ L_\text{R}\cos{ \theta }\\
=( L_\text{I}-  L_\text{T} + L_\text{R})\cos{\theta}\\
=( \eta_\text{I}-  T\eta_\text{T} + R\eta_\text{R})I\cos{\theta}
\end{aligned}
\end{equation}

\indent
$L _\text{I}$,  $L_\text{T}$, $L_\text{R}$ and $\eta_\text{I}$,  $\eta_\text{T}$, $\eta_\text{R}$ are angular momenta and ellipticity for incident, transmitted and reflected light beams, respectively. $\text{I}$  and $\theta$ are intensity and incident angle of the incident beam.\\
\indent
 {\small QWP} and {\small HWP} angle dependence were calculated by Jones calculus. For several  {\small QWP} angles, {\small HWP} angle dependence were calculated. Ellipticity was estimated by

\begin{equation}
{\bf{E_{0}}}=^{t}(1,0) \qquad\text{,}\qquad \bf{E_\text{I}}(\alpha, \beta)=\bf{H( \beta)Q(\alpha)E_{0}}.
\end{equation}

\begin{equation}
\bf{E}_\text{R}(\alpha, \beta)= {F}_\text{R}(\theta){E}_\text{I}(\alpha,\beta)      \qquad\text{,}\qquad      {E}_\text{T}(\alpha, \beta)={F}_\text{T}(\theta){E}_\text{I}(\alpha, \beta).
\end{equation}

{\begin{eqnarray}
\bf{Q}(\alpha) &=&
\begin{pmatrix}
\cos{\alpha}&-\sin{\alpha} \\
\sin{\alpha}&\cos{\alpha}
\end{pmatrix}
\begin{pmatrix}
i&0 \\
0&1
\end{pmatrix}
\begin{pmatrix}
\cos{\alpha}&\sin{\alpha} \\
-\sin{\alpha}&\cos{\alpha}
\end{pmatrix}
\end{eqnarray}

\begin{eqnarray}
\bf{H}(\beta)&=&
\begin{pmatrix}
\cos{\beta}&-\sin{\beta} \\
\sin{\beta}&\cos{\beta}
\end{pmatrix}
\begin{pmatrix}
exp{(i\delta_{\small HWP})}&0 \\
0&1
\end{pmatrix}
\begin{pmatrix}
\cos{\beta}&\sin{\beta} \\
-\sin{\beta}&\cos{\beta}
\end{pmatrix}
\end{eqnarray}

\begin{eqnarray}
\bf{F}_\text{R}&=&
\begin{pmatrix}
r_\text{p}e^{i\delta_{r}}&0\\
0&r_\text{s}
\end{pmatrix}
\end{eqnarray}

and 

\begin{eqnarray}
\bf{F}_\text{T}&=&
\begin{pmatrix}
t_\text{p}e^{i\delta_{t}}&0\\
0&t_\text{s}
\end{pmatrix}
\end{eqnarray}

\begin{equation}
\delta_{r}=\delta_{r\text{p}}-\delta_{r\text{s}} \qquad\qquad\qquad\qquad \delta_{t}=\delta_{t\text{p}}-\delta_{t\text{s}}
\end{equation}

\begin{equation}
\eta_\text{J}=\frac{i\bf{E}_\text{J}\times\bf{E}^{*}_\text{J}}{{|\bf{E}_\text{J}|}^2},     \qquad \text{ J}=\text{I}, \text{ R},  \text{ T}
\end{equation}

\indent
For a proper {\small HWP}, $\delta_{\small HWP}=\pi$. Complex Fresnel coefficients of our sample were determined from a separate experiment at 530 nm using a conventional ellipsometry technique. The results are:\\

\begin{equation}
r_\text{p}={(0.17)}^{1/2} \qquad\text{,}\qquad r_\text{s}={(0.33)}^{1/2}\qquad\text{,}\qquad \delta_{r}=0.77\pi
\end{equation}

\begin{equation}
t_\text{p}={(0.29)}^{1/2}\qquad\text{,}\qquad t_\text{s}={(0.16)}^{1/2} \qquad\text{,}\qquad \delta_{t}=0.02\pi
\end{equation}

Results in Fig. \ref{All} shows calculated $\Delta L_\text{Sample}$ as a function of the angles of the rotating {\small HWP} for each polarization. As we see in Fig. \ref{All}, calculation of the $\Delta L_\text{Sample}$ as a function of the angle of {\small HWP} with $\delta_{\small HWP}=0.9\pi$ reproduces the measurement results fairly well. This excellent agreement suggests that {\small TPIV} is indeed ascribed to the angular momentum transfer. A separate experiment of extinction ratio of the {\small HWP} turned out to be consistent with this phase delay.

\begin{figure}[t]
\footnotesize
\includegraphics[trim = 0mm 55mm 5mm 50mm, clip, width=\linewidth]{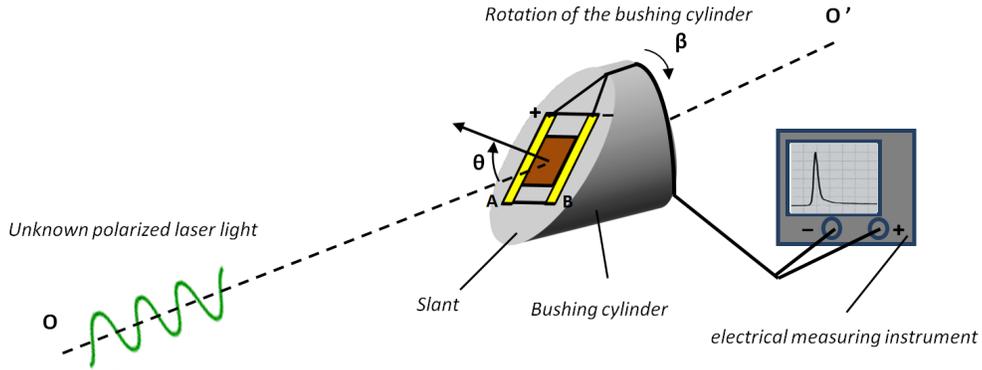} 

\caption{\label{IMG2} Bushing cylinder which rotates around axis {$\small OO'$} that coincides with propagation direction of the unknown light. For an unknown polarized light (green wave) by rotating the bushing, relative orientation of the incident light polarization to the electrodes is changed, for which generated {\small TPIV} is measured.}
\end{figure}
\section{Proposal} 

{\small TPIV} patterns for each polarized input light can be used in design of a polarimeter to determine ellipticity of any unknown polarized light. Although we used an achromatic {\small HWP} to rotate the polarization direction in our present experiment, we can rotate the sample instead. In this case we need not rely on the very accurate achromatic wave plate, which is a great advantage of the device. For this purpose we employ a design devised to determine linear polarization direction \cite{Mikheev2012B}. This polarimeter consists of a bushing cylinder and a slant on it for mounting the thin film with two electrode. They are connected to an electrical measuring instrument and allow us to monitor the generated photovoltage (Fig. \ref{IMG2}). \\
\indent
As {\small TPIV} is proportional to the angular momentum transfer rather than the angular momentum carried by the incident beam, {\small TPIV} itself is not the measure of the ellipticity of the incident beam. As the bushing cylinder rotates, incident plane rotates while the two electrodes measure {\small TPIV} for different polarization orientation. It corresponds to having a fixed film and light with rotating orientation apply on the film. The $\beta$ dependence shows specific pattern depending on the ellipticity carried by the incident beam. By comparing the pattern to the calculation, it is possible to determine the ellipticity of the incident beam. The ratio of average and amplitude gives an index that is related to the ellipticity. In order to make a quantitative determination of ellipticity, however, calibration due to wavelength dependence of  Fresnel coefficients is necessary. \\
\indent
In contrast to polarization analyzer designed in \cite{Mikheev2012B} our polarimeter can be used to determine any linearly, circularly and elliptically polarized unknown light in visible frequencies. Fabrication method of the thin film is simple compared to other materials used to fabricate such analyzers, for example,  single walled carbon nanotube and nanographite.\\
\indent
Alll electric semiconductor based detectors for Stokes parameter including helicity has been proposed since 2008 \cite{Ganichev2008}. To characterize the radiation helicity several mechanisms has been implemented including the circular photogalvanic effect in semiconductor quantum wells  \cite{Danilov2009}. Dyakonov/Shur mechanism of THz detection in HEMT structures \cite{Drexler2012} and ratchet effect in low dimensional semiconductor structures with asymmetric lateral superlattices   \cite{Faltermeier2015}. If we design a metamaterial so that it satisfies the symmetry requirement, it may be possible to achieve all electric detector for polarization state.
\section{Conclusion}

In summary we have experimentally investigated transverse photo-induced voltage {\small TPIV} in 25 nm thick gold film with random nanoholes prepared by {\small NSL} technique. While {\small TPIV} is not generated by s- and p- polarized incident light, linearly, circularly and elliptically polarized light can generate {\small TPIV}. Sign and amplitudes of the voltage dependent on the sense of polarization, ellipticity and degree of polarization in circularly, elliptically and linearly polarized light. Calculation based on angular momentum transfer shows that transferred angular momentum from photon to the film is responsible for this voltage. Based on this mechanism we have proposed a polarimeter that determines ellipticity.\\

\end{document}